# Interactions Between Artificial Intelligence and Digital Public Infrastructure: Concepts, Benefits, and Challenges

Sarosh Nagar
University College London
Institute for Innovation and Public Purpose
ucbvsnn@ucl.ac.uk

David Eaves
University College London
Institute for Innovation and Public Purpose
d.eaves@ucl.ac.uk

Abstract

Artificial intelligence (AI) and digital public infrastructure (DPI) are two technological developments that have taken center stage in global policy discourse. Yet, to date, there has been relatively little discussion about how AI and DPI can mutually enhance the public value provided by each other. Therefore, in this paper, we describe both the opportunities and challenges under which AI and DPI can interact for mutual benefit. First, we define both AI and DPI to provide clarity and help policymakers distinguish between these two technological developments. Second, we provide empirical evidence for how AI, a general-purpose technology, can integrate into many DPI systems, aiding DPI function in use cases like language localization via machine translation (MT), personalized service delivery via recommender systems, and more. Third, we catalog how DPI can act as a "foundation" for creating more advanced AI systems by improving both the quantity and quality of training data available. Fourth, we discuss the challenges of integrating AI and DPI, including high inference costs for advanced AI models, interoperability challenges with legacy software, concerns about induced bias in AI systems, and privacy challenges related to DPI. We conclude with key takeaways for how policymakers can work to enhance the positive interactions of AI and DPI.

**Policy Significance Statement**
This commentary is significant to policymakers for four reasons. First, it provides clear, robust definitions of both AI and DPI to improve policymakers' precision on these topics. Second, for policymakers whose governments are implementing DPI, this paper provides clear use cases for the integration of advanced AI into DPI systems. Third, this paper outlines a long-term vision for how DPI can serve as the foundation of better frontier AI models through improved data governance. Fourth, this paper provides actionable policy recommendations for governments seeking to incentivize the positive interaction of the two technologies.

## 1. Introduction

Two technological developments are gaining significant traction worldwide. One is artificial intelligence (AI), and in particular, machine learning (ML) and deep learning (DL).[i] Since the advent of OpenAI's ChatGPT, AI has become a fixture of global fora like the United Nations Summit of the Future and a central domain over which policymakers are increasingly focused.[ii] The other major technological trend is digital public infrastructure (DPI), which makes public key digital systems to enable new innovations and facilitate broader digital transformation.[iii] DPI technically takes the form of a platform layer of software that connects physical devices like routers to an application layer with various digital solutions in areas like payment, and governments, in turn, use DPI to deliver public services, foster economic development through reduced expenditure, lower transaction costs, enable competition, assert digital sovereignty, and more. Several countries, like India, have already established widespread digital identification and payment systems as part of their DPI offerings.[iv,v]

Despite global interest in these two technological developments, however, there has been little discussion of how they might interact. Therefore, in this paper, we aim to fill this gap in the literature. First, we provide clear definitions of AI and DPI for policymakers, highlighting the distinction between both technological developments, their use cases, and more. Second, we review how these technologies could interact, exploring both the potential of AI to enhance the public value created by DPI and the role of DPI in acting as a 'foundation' for better AI development in the long run. We also review the challenges associated with facilitating these interactions and conclude with recommendations that policymakers and firms can use to enhance the positive use of AI and DPI.

## 2. Defining AI and DPI
### 2.1. Defining artificial intelligence (AI)

Artificial intelligence (AI) emerged from developments in computer science, neuroscience, electrical engineering, and other fields, and as coined by Stanford Professor John McCarthy in 1955, is "the science of making intelligent machines".[vi] Most modern AI tools with which policymakers, interact, in turn emerge from a subfield within computer science called machine learning (ML), and in particular, the specific subfield of deep learning (DL), which relies on a type of system called a neural network to process information.[vii] In turn, one specific design or architecture for these neural network architecture, called a transformer, forms the basis for modern AI systems like ChatGPT, as transformers take in large amounts of input data and processes them to produce a given response.[viii] Beyond these large language models (LLMs), however, there are many other types of AI models, like diffusion models, which form the basis of many image generation and drug discovery system.[ix] AI, in turn, has a wide range of potential benefits it can offer to scientific research, economic growth, and more.

From the lens of a policymaker, it is worth recognizing that AI is a general-purpose technology (GPT), like electricity before it.[x,xi] This designation means that AI is a technology with many broad applications across society, giving policymakers a vast area of potential areas in which the technology can be used to increase public value, but also giving these groups an important challenge to correctly identify and design integrations of the technology into existing digital or physical solution.[xii] Furthermore, the broad nature of general-purpose technology (GPT) it is often difficult to discretely classify specific use cases or categories of the technology, much as it is difficult to neatly systematize all potential applications of electricity. The result is a challenge that becomes relevant when discussing applications of AI for DPI, as the potential applications are vast but are difficult to discretely categorize.

### 2.2. Defining digital public infrastructure (DPI)

As opposed to AI, digital public infrastructure (DPI) is not a singular technology but rather a specific application of numerous digital technologies. As stated earlier, DPI, in a narrow technical sense, refers to sets of software that make up an intermediate platform layer between physical technologies, like data centers, and the application layer.[xiii] More broadly, however, this software is designed to act akin to a core piece of infrastructure, enabling day-to-day functions like digital payment, identification, and more, leading to the use of the terms "digital" and "infrastructure" in DPI.[xiv] In turn, the "public" in digital public infrastructure comes from the fact that for much of this infrastructure to work, government oversight in often necessary in certain domains, such as in ensuring certain security standards.[xv] DPI, in turn, has transformative potential to lay the foundations of what a modern digital state might look like.

While the types of systems which compose DPI vary, three particular types are common. First, there are digital identity systems, which refer to a broader software system that enables individuals to establish and verify their digital identities[xvi]. One example of a digital identity system is India's Aadhar, which provides digital identification that citizens can use for identify verification, setting up bank accounts, and more.[xvii] The second form is digital payment, which refers to digital solutions that can facilitate rapid transactions between two financial actors.[xviii] One famous example of a digital payment system is the Central Bank of Brazil's Pix system, which facilitates instant digital payments between two parties in the country.[xix] The third common type of digital system described as DPI is data exchange systems (DES), which manage the secure flow of information between public and private actors and across multiple platforms.[xx] Estonia's X-Road is one example of a data exchange system that enables secure data transfers across multiple systems, allowing for rapid sharing of data from government entities to the private sector, within state agencies, and more.[xxi] Of course, other forms of DPI systems can exist, but these three systems are often ubiquitous in the most successful cases of DPI, and hence are particularly worth discussing in the context of this paper.

## 3. Interactions between AI and DPI

With definitions of AI and DPI, we can begin analyzing where the two technologies can interact with each other. The integration of AI into DPI can increase DPI's usefulness and accessibility across a wide variety of domains, given AI's general-purpose nature, in turn enhancing the public value that DPI more broadly creates for society. Meanwhile, the data governance and input from DPI can provide a solid foundation for the development of robustly secure and advanced frontier AI systems. There are a number of challenges, however, that policymakers face in realizing these benefits, requiring further attention and action to overcome these issues.

### 3.1. AI as an enhancer of DPI

AI is a general-purpose technology while DPI is a specific application of digital technologies into a multi-purpose platform layer.[xxii,xxiii] As a result, AI's general-purpose nature means it can enact as a broader enhancer of DPI in a wide variety of use cases, as the technology can be integrated into a number of different digital systems to improve the public value offered by those systems. Simultaneously, this property also means it is difficult to systematize all the use cases for AI in DPI, much as it would be difficult to systematize all use cases of any general-purpose technology in society. Thus, in this section, we describe a few empirically validated examples of where governments can use AI to improve DPI's social benefits.

One empirically validated use case for AI in enhancing the public value created by DPI is language localization in service delivery. Across several major economies, linguistic diversity, while certainly valuable culturally, can sometimes impose transaction costs on the economy and make it more difficult for governments to provide public services to the whole country as well as complicate interactions between individuals.[xxiv] As a result, DPI systems in many societies must be multilingual to reach and engage these linguistic minorities. In turn, modern large language models (LLMs) can be a valuable tool in facilitating multilingual DPI. LLMs can facilitate higher-quality machine translation (MT) to individuals who speak different languages.[xxv] As a result, integrating LLMs into DPI systems can allow individuals speaking various languages to interact easily with digital payment and identity systems, reducing these transaction costs and facilitating greater public service uptake in these communities.[xxvi,xxvii] One empirical example of this phenomenon comes from India, which is building Bhashini, an AI-led language translation system that aims to integrate with India's other DPI systems.[xxviii] Bhashini utilizes natural language processing (NLP), a machine learning technique most commonly used with text, to perform rapid translations between Indic languages.[xxix] In order to improve the system's translations, especially in languages where prior data may be limited, Bhashini utilizes a nationwide network of crowdsourced collaborators to transcribe content across Indic languages and validate crowdsourced translations made by others.[xxx] As a result, this feedback across the national network is designed to improve the quality of Bhashini's translations, and the Indian governments eventually hopes to deploy this high-power system across India's other DPI systems to enable the country's linguistic minorities to gain greater access to digital public services.[xxxi]

Of course, other use cases exist beyond language delivery as well. For example, Singapore is using AI to combat fraud in DPI, as Singapore's national DPI, Singpass, recently began using a machine learning to authenticate the users of the DPI system, helping improve citizens' trust in the system, which can facilitate broader uptake.[xxxii] Denmark, meanwhile, is using AI to streamline public service delivery, recently launching Muni, a chatbot that helps users across 37 Danish municipalities access digital public services.[xxxiii] Furthermore, some have suggested that integrating AI into DPI can also enable the personalization of DPI through "recommender systems," which are a class of algorithm that can personalize content recommendations to individuals.[xxxiv] These recommender systems could be deployed in DPI to personalize the digital services individuals are offered, ensuring that broader public services more effectively meet their needs. While empirical examples with DPI and recommender systems are currently rare, countries like Singapore have begun experimenting with using AI to personalize interventions in fields like healthcare, and it is conceivable similar efforts could occur with other DPI services.[xxxv]

The examples of language localization, fraud detection, and personalization, of course, are only a few potential use cases of AI in DPI — and there are likely many others that have not been publicly disclosed or which still remain in the ideation stage. Nevertheless, they highlight how AI's general-purpose nature enables it to enhance the use of DPI in fundamentally transformative ways. It is highly likely that, in turn, as these new applications for AI/ML emerge in DPI systems, further analysis will be needed to catalog the wide variety of positive uses that may emerge from these developments.

3.2. DPI as a foundation for better frontier AI

While AI can enhance many aspects of DPI, DPI is not to be left out of these discussions. Indeed, through the data collected and curated by DPI, DPI can play a transformative role as a foundation for better AI systems, much as better roads and general physical infrastructure build the foundation for advanced vehicles. At the center of this foundation element is data — a key input to modern AI systems, which DPI has the unique potential to greatly improve. In particular, DPI can improve the supply and quality of high-training and post-training data for advanced AI models, aid in the incorporation of unwritten and traditional knowledge (TK) into AI's training datasets, and more.

First, DPI can enhance the development of advanced AI by improving the supply of high-quality data. Currently, there are substantial fears that large language models will consume all human-generated data by 2028, with solutions like AI-generated synthetic data being imperfect substitutes due to risks that extensively training AI on synthetic data can induce the "collapse" of AI models or cause significant declines in the diversity of outputs from a model.[xxxvi, xxxvii] Therefore, there is a vital need to acquire more data to enable advancements in AI systems. DPI, however, represents a values-consistent solution to this problem. DPI systems, like digital payment, collect large amounts of data from citizens, but only with citizens' consent.[xxxviii] As a result, that vast quantity of consent-based data collected can be used to enhance the training and post-training data of frontier AI models. As a result, large quantities of human-generated but ethically collected data from DPI systems may be highly valuable in bypassing the 'data wall' in AI development to enable more performant frontier AI systems.

Indeed, evidence is already emerging that the amount of data DPI could produce could serve this purpose, as seen in India. India uses a digital identification system called Aadhar as part of its broader DPI offerings, and to date, Aadhar has nearly 1.38 billion users.[xxxix,xl] Seeing a large amount of consent-based data this collects, the Indian government, in turn, has placed this massive volume of data into public datasets, which other Indian AI startups can use to develop their own AI/ML systems.[xli,xlii] This unique arrangement, therefore, could be described as an informal version of 'industrial policy' for AI, as it gives India's private sector access to a tremendous amount of data that they can use to train higher-quality frontier AI systems. In turn, the value of this data will only grow over time because as the quantity of human-generated data continues to decline, states that can access more of it will be uniquely positioned to lead the world on AI. Thus, we see how DPI can potentially offer a consent-based solution to constraints about data supply in the future of frontier model development.

However, it is not just in the quantity of data where DPI is powerful — but also in quality. Much of the improvement of frontier AI models' performance beyond the level of OpenAI's GPT-4 has been due to increases enabled by high-quality post-training data, but once again, data constraints may limit how much high-quality data is present.[xliii] The result means that as the supply of high-quality data declines, the gains which new AI models can reap from high-quality post-training data will decline, in turn leading to a slower rate of growth in the performance of frontier AI systems. However, the data produced by DPI may be able to rectify this problem. Much of the data on the internet is unstructured, non-standardized, and, in some cases, may be of low quality, rendering it challenging to train models on these datasets and generate significant performance gains from such data. However, DPI systems have a standardized format for the datasets they generate, as seen in the case of Mauritius, whose DPI systems helped create standardized protocols for data exchange and storage.[xliv,xlv] The result enables the creation of structured, high-quality datasets, which, if used in model post-training, could potentially lead to significant gains in model performance over time.

Beyond data standardization, however, DPI can also tackle another issue affecting dataset quality — bias. Certain historically marginalized populations are often underrepresented in AI datasets, resulting in algorithmic bias that means AI models can make harmful or inaccurate decisions when engaging with these communities.[xlvi] Furthermore, certain types of knowledge, like the traditional knowledge (TK) of Indigenous populations about medicinal formulas, rarely make it into AI training datasets.[xlvii] Such data, however, can be valuable for training general-purpose AI (GPAI) models, especially those used in scientific fields, as exemplified by the origins of the malarial drug artemisinin, which came from traditional Chinese medicinal knowledge.[xlviii] Therefore, there is a broad need to collect high-quality data with reduced bias against marginalized populations and better incorporation of unwritten traditional knowledge into their systems. DPI systems, however, represent another solution to this data bias challenge. In theory, DPI platforms are universal platforms that will be used by members of marginalized and Indigenous communities. As a result, if consent-based data is provided by these communities via DPI, the result means DPI will capture the unwritten knowledge or information from these communities, as well as, more broadly, ensure more data emerges from historically marginalized groups. When model developers then train their systems on these datasets, the result means that the training data will be more representative, capturing TK and other information often left out of training datasets. As a result, when AI models are deployed for fields like scientific research, this broader knowledge can be leveraged by the system to produce more innovative breakthroughs.

There may, of course, be other ways in which DPI can be the foundation for advanced AI systems, but data governance arguably remains the most immediate way. DPI can serve as the bedrock for high-quality, consent-based data for better frontier AI systems, which, in turn, will lead to greater improvements in model performance gain over time.

### 4. Challenges

The integration of AI and DPI offers many potential benefits, but it is not like reaping these benefits will be a seamless process either. In reality, significant technical and political barriers also create several challenges in realizing the benefits of AI and DPI's interactions. These challenges are varied: some are technical, while others are considerations of the policymaking processes, and others are still driven by important values-based concerns. Together, however, these challenges need to be addressed to bring the broader benefits of AI and DPI together.

Most immediately, there are challenges hindering AI's ability to integrate into DPI. One chief concern has to do with the inference costs associated with running an AI model.[xlix] While model inference is cheaper than model training, the fact that DPI is deployed on a national scale to hundreds of millions or even billions of citizens means that integrating certain AI features into DPI may force governments to pay high inference costs given the volume of individuals using frontier models through DPI.[l] This problem may be particularly acute since, with the advent of technical improvements which increase test-time compute, like best-of-n sampling or chain of thought reasoning, as displayed in OpenAI's o1 model.[li,lii] These types of model improvements would increase inference costs per query, which yield substantially higher costs if an entire country is accessing these inference cost-intensive AI systems as part of DPI.[liii, liv] Furthermore, these higher inference costs can also strain government finances and exceed the data center and power limits of hyperscalers if enough inference calls are made. Therefore, when integrating AI into DPI, governments should be careful to identify which specific AI integrations in DPI can be done by smaller models with cheaper inference and which ones require more computationally expensive systems.

A second problem further hindering the use of AI in DPI is related to interoperability. Many governments worldwide use a mix of modern and legacy software systems, but the software underlying advanced AI may not be interoperable with the legacy software or even some modern software used in DPI or other government applications. Indeed, the United Kingdom's National Audit Office flagged these interoperability challenges as a central barrier to AI adoption in the government more broadly.[lv] The World Bank highlights a similar set of challenges across many Global South nations as well.[lvi] The result is that such interoperability barriers may hinder the integration of AI systems and DPI more broadly, preventing the creation of the public value associated with their interaction. To overcome such challenges, governments may need to redesign significant portions of the codebases underlying their DPI systems to ensure interoperability with advanced AI/ML systems, which can take time and may be costly for some states.

Not to be outdone, efforts to use DPI as a foundation for more powerful AI development also face its own hurdles as well. One immediate problem is that many of the benefits DPI can offer for AI development — such as improvements in the quality and quantity of data — require an inclusive and effective implementation of DPI, or else risk becoming a double-edged sword.[lvii] For example, in order for data from DPI to help combat algorithmic bias, DPI systems must be adopted by members of marginalized communities, who may face systemic disparities that limit their ability to use these digital tools or may be hesitant to use government platforms due to legacies of marginalization.[lviii] The result is that the traditional knowledge gains or other benefits from incorporating these communities into datasets may be lost absent efforts to ensure DPI reaches all parts of the population. As a result, prior to using data from DPI for public datasets, governments should be careful to audit DPI data to ensure the data comes from a representative sample of their citizens — otherwise, DPI datasets may only exacerbate algorithmic bias in the long run.

Second, it is worth highlighting that the use of DPI data for public datasets comes with important ethical and political considerations. If governments intend to follow India's lead in creating public AI training datasets from DPI data, for example, it is vital that all data is collected with individuals' informed consent before its collection — otherwise, DPI risks infringing on important individual rights for privacy.[lix] This data must also be stored in a secure place, with necessary cybersecurity and localization protections — otherwise, the sensitive data of millions of citizens may face elevated cybersecurity risks due to poor policy choices.[lx] In addition, there may be adoption challenges related to low levels of trust in government in some parts of the world, as some countries' populations may not wish to use DPI at the levels necessary to truly empower AI systems. These examples are only some of the numerous ethical and values-based considerations that must be factored in when governments are deciding whether to use DPI to enhance the quality of AI tools.[lxi]

More broadly, it is important that policymakers distinguish between AI and DPI in policy discussions. Broadly grouping these two technological developments into the same national digital strategy documents can cause confusion, especially since policy debates that seem similar across both AI and DPI can be quite different. For example, discussions of "open source" in the DPI context refer to open-sourcing all the code associated with a given DPI system, like data exchange, while there is debate in the AI community over what this term even means, as some accuse models which claim to be "open source" as not providing enough information to truly be open source.[lxii] As a result, attempting to address open-source for both DPI and AI under the same basket or within the same document risks muddling the costs and benefits of each approach together.[lxiii] The result creates unclear policy discourse that can hamper domestic AI development and use of AI for the public good.

Once again, however, these challenges are only a subset of those that may challenge the positive integration of AI and DPI for the public good. It is vital that governments, firms, and other actors move to combat these challenges to ensure the broader benefits of AI and DPI.

## 6. Conclusion

AI and DPI are some of the most transformative technological developments of the 21st century. Both technologies offer huge potential to transform economies and reshape governments for the public good. AI is a general-purpose technology capable of revolutionizing nearly all sectors of society, while DPI creates a foundational layer that can support the well-being of many millions and create broader public value for all of society. In this paper, we highlight how AI can enhance DPI through personalization, language localization, and a broad range of other integrations in various DPI systems. Meanwhile, we also discuss how DPI can act as the foundation for better frontier AI models through improved data governance, debiasing, and standardization. Of course, reaping these benefits will require overcoming significant technical, political, and ethical hurdles related to interoperability, inference costs, and political considerations. However, together, overcoming them will be vital to ensure a positive digital future for humanity.

**Word Count: 3920**

**Acknowledgments.** The author is grateful for the comments provided by the IIPP Digital Team, especially Beatriz Vasconcellos, Kassim Vera, Navanit Kumar, and Krisstina Rao.
**Funding statement.** None.
**Competing interests.** None.
**Data availability statement.** None.
**Author contributions.** Conceptualization: S.N. Writing original draft: S.N. Critical revisions: S.N.; D.E.. All authors approved the final submitted draft.

# References

[i] Ekman, M. (2021). *Learning deep learning: Theory and practice of neural networks, computer vision, natural language processing, and transformers using TensorFlow*. Addison-Wesley Professional.

[ii] Belhaj, F. (2024). The UN-led Summit for the Future: Multilateral Solutions for a Better Tomorrow.

[iii] Eaves, D., Mazzucato, M., & Vasconcellos, B. (2024). Digital public infrastructure and public value: What is 'public' about DPI?.

[iv] Rodriguez, S, Price, & Rodriguez, A. (2023). Reimagining our Digital Future. *New America*.

[v] Eaves, D., Mazzucato, M., & Vasconcellos, B. (2024). Digital public infrastructure and public value: What is 'public' about DPI?.

[vi] Manning, C. (2020). Artificial Intelligence Definitions. *Stanford University Human-Centered Artificial Intelligence (HCAI)*. https://hai.stanford.edu/sites/default/files/2020-09/AI-Definitions-HAI.pdf

[vii] Ekman, M. (2021). *Learning deep learning: Theory and practice of neural networks, computer vision, natural language processing, and transformers using TensorFlow*. Addison-Wesley Professional.

[viii] Ekman, M. (2021). *Learning deep learning: Theory and practice of neural networks, computer vision, natural language processing, and transformers using TensorFlow*. Addison-Wesley Professional.

[ix] Alakhdar, A., Poczos, B., & Washburn, N. (2024). Diffusion Models in $\textit{De Novo}$ Drug Design. *arXiv preprint arXiv:2406.08511*.

[x] Crafts, N. (2021). Artificial intelligence as a general-purpose technology: an historical perspective.

[xi] Ekman, M. (2021). *Learning deep learning: Theory and practice of neural networks, computer vision, natural language processing, and transformers using TensorFlow*. Addison-Wesley Professional.

[xii] Ekman, M. (2021). *Learning deep learning: Theory and practice of neural networks, computer vision, natural language processing, and transformers using TensorFlow*. Addison-Wesley Professional.

[xiii] Rodriguez, S, Price, & Rodriguez, A. (2023). Reimagining our Digital Future. *New America*.

[xiv] Eaves, D., Mazzucato, M., & Vasconcellos, B. (2024). Digital public infrastructure and public value: What is 'public' about DPI?.

[xv] Eaves, D., Mazzucato, M., & Vasconcellos, B. (2024). Digital public infrastructure and public value: What is 'public' about DPI?.

[xvi] Eaves, D., Mazzucato, M., & Vasconcellos, B. (2024). Digital public infrastructure and public value: What is 'public' about DPI?.

[xvii] DPI Map (2024). dpimap.org

[xviii] Eaves, D., Mazzucato, M., & Vasconcellos, B. (2024). Digital public infrastructure and public value: What is 'public' about DPI?.

[xix] DPI Map (2024). dpimap.org

[xx] Eaves, D., Mazzucato, M., & Vasconcellos, B. (2024). Digital public infrastructure and public value: What is 'public' about DPI?.

[xxi] DPI Map (2024). dpimap.org

[xxii] Renda, A. (2019). *Artificial Intelligence. Ethics, governance and policy challenges*. CEPS Centre for European Policy Studies.

[xxiii] Eaves, D., Mazzucato, M., & Vasconcellos, B. (2024). Digital public infrastructure and public value: What is 'public' about DPI?.

[xxiv] Gurevich, T., Herman, P., Toubal, F., & Yotov, Y. V. (2021). One nation, one language? domestic language diversity, trade and welfare.

[xxv] Wang, L., Lyu, C., Ji, T., Zhang, Z., Yu, D., Shi, S., & Tu, Z. (2023). Document-level machine translation with large language models. *arXiv preprint arXiv:2304.02210*.

[xxvi] Alonso, C., Bhojwani, T., Hanedar, E., Prihardini, D., Uña, G., & Zhabska, K. (2023). *Stacking up the benefits: Lessons from India's digital journey*. International Monetary Fund.

xxvii Wang, L., Lyu, C., Ji, T., Zhang, Z., Yu, D., Shi, S., & Tu, Z. (2023). Document-level machine translation with large language models. *arXiv preprint arXiv:2304.02210*.

xxviii Shubham, S. (2023). Explained: What is Bhashini, the real-time translation tool that PM Narendra Modi used. *Times of India*.

xxix xxix Shubham, S. (2023). Explained: What is Bhashini, the real-time translation tool that PM Narendra Modi used. *Times of India*.

xxx Alonso, C., Bhojwani, T., Hanedar, E., Prihardini, D., Uña, G., & Zhabska, K. (2023). *Stacking up the benefits: Lessons from India's digital journey*. International Monetary Fund.

xxxi Ministry of Electronics and Information Technology. n.d. National Language Translation Mission (NLTM) Whitepaper.

xxxii Theodorou, Y., Cowie, C., & Charlwood, R. (2024). Creating Value for Users and Governments: How AI Can Enhance Digital ID Solutions. *Tony Blair Institute for Global Change*.

xxxiii EuroCities. (2023). Accessing Danish digital public services with Muni. *Eurocities*. https://eurocities.eu/latest/accessing-danish-digital-public-services-with-muni/

xxxiv Roy, D., Dutta, M. A systematic review and research perspective on recommender systems. *J Big Data* **9**, 59 (2022). https://doi.org/10.1186/s40537-022-00592-5

xxxv Theodorou, Y., Cowie, C., & Charlwood, R. (2024). Creating Value for Users and Governments: How AI Can Enhance Digital ID Solutions. *Tony Blair Institute for Global Change*.

xxxvi Pablo Villalobos, Anson Ho, Jaime Sevilla, Tamay Besiroglu, Lennart Heim, and Marius Hobbhahn. 'Will we run out of data? Limits of LLM scaling based on human-generated data'. *ArXiv [cs.LG]*, 2024. arXiv. https://arxiv.org/abs/2211.04325.

xxxvii Seddik, M. E. A., Chen, S. W., Hayou, S., Youssef, P., & Debbah, M. (2024). How bad is training on synthetic data? a statistical analysis of language model collapse. *arXiv preprint arXiv:2404.05090*.

xxxviii Eaves, D., Mazzucato, M., & Vasconcellos, B. (2024). Digital public infrastructure and public value: What is 'public'about DPI?.

xxxix Press Information Bureau, Government of India. (2024). Government of India taking measures to enhance the reach of Indian Digital Public Infrastructure. *Government of India*.

xl Rajvanshi, A. (2024). India is Emerging as a Key Player in the Global AI Race. *Time*.

xli Press Information Bureau, Government of India. (2024). Government of India taking measures to enhance the reach of Indian Digital Public Infrastructure. *Government of India*.

xlii Rajvanshi, A. (2024). India is Emerging as a Key Player in the Global AI Race. *Time*.

xliii Wang, A., Schneider, J, & Ottinger, L. (2024). Scale's Alex Wang on the US-China AI Race. *ChinaTalk*.

xliv Bandura, R., McLean, M., Sultan, S. (2023). Unpacking the Concept of Digital Public Infrastructure and its Importance for Global Development. *Center for Strategic and International Studies.* https://www.csis.org/analysis/unpacking-concept-digital-public-infrastructure-and-its-importance-global-development

xlv Schwabe, D., Becker, K., Seyferth, M., Klaß, A., & Schaeffter, T. (2024). The METRIC-framework for assessing data quality for trustworthy AI in medicine: a systematic review. *NPJ Digital Medicine*, *7*(1), 203.

xlvi Gichoya, J. W., Thomas, K., Celi, L. A., Safdar, N., Banerjee, I., Banja, J. D., ... & Purkayastha, S. (2023). AI pitfalls and what not to do: mitigating bias in AI. *The British Journal of Radiology*, *96*(1150), 20230023.

xlvii Jones, P. L., Mahelona, K., Duncan, S., & Leoni, G. (2023). Kia tangata whenua: Artificial intelligence that grows from the land and people.

xlviii Czechowski, T., Weathers, P. J., Brodelius, P. E., Brown, G. D., & Graham, I. A. (2020). Artemisinin—from traditional Chinese medicine to artemisinin combination therapies; four decades of research on the biochemistry, physiology, and breeding of Artemisia annua. *Frontiers in Plant Science*, *11*, 594565.

[xlix] Luccioni, S., Jernite, Y., & Strubell, E. (2024, June). Power hungry processing: Watts driving the cost of AI deployment?. In *The 2024 ACM Conference on Fairness, Accountability, and Transparency* (pp. 85-99).

[l] Luccioni, S., Jernite, Y., & Strubell, E. (2024, June). Power hungry processing: Watts driving the cost of AI deployment?. In *The 2024 ACM Conference on Fairness, Accountability, and Transparency* (pp. 85-99).

[li] Snell, C., Lee, J., Xu, K., & Kumar, A. (2024). Scaling llm test-time compute optimally can be more effective than scaling model parameters. *arXiv preprint arXiv:2408.03314*.

[lii] Wei, J., Wang, X., Schuurmans, D., Bosma, M., Xia, F., Chi, E., ... & Zhou, D. (2022). Chain-of-thought prompting elicits reasoning in large language models. *Advances in neural information processing systems*, *35*, 24824-24837.

[liii] Snell, C., Lee, J., Xu, K., & Kumar, A. (2024). Scaling llm test-time compute optimally can be more effective than scaling model parameters. *arXiv preprint arXiv:2408.03314*.

[liv] Wei, J., Wang, X., Schuurmans, D., Bosma, M., Xia, F., Chi, E., ... & Zhou, D. (2022). Chain-of-thought prompting elicits reasoning in large language models. *Advances in neural information processing systems*, *35*, 24824-24837.

[lv] National Audit Office. (2024). Use of artificial intelligence in government. *UK Department for Science, Innovation, & Technology.*

[lvi] World Bank. (2021). Artificial Intelligence in the Public Sector: Summary Note.

[lvii] Schwabe, D., Becker, K., Seyferth, M., Klaß, A., & Schaeffter, T. (2024). The METRIC-framework for assessing data quality for trustworthy AI in medicine: a systematic review. *NPJ Digital Medicine*, *7*(1), 203.

[lviii] Saeed, S. A., & Masters, R. M. (2021). Disparities in health care and the digital divide. *Current psychiatry reports*, *23*, 1-6.

[lix] Miller, S., & Bossomaier, T. (2024). *Cybersecurity, Ethics, and Collective Responsibility*. Oxford University Press.

[lx] Rafiq, F., Awan, M. J., Yasin, A., Nobanee, H., Zain, A. M., & Bahaj, S. A. (2022). Privacy prevention of big data applications: A systematic literature review. *SAGE Open*, *12*(2), 21582440221096445.

[lxi] Miller, S., & Bossomaier, T. (2024). *Cybersecurity, Ethics, and Collective Responsibility*. Oxford University Press.

[lxii] United Nations Development Programme. (2023). The Human and Economic Impact of Digital Public Infrastructure. *United Nations Development Programme.*

[lxiii] Davies, P. (2024). Open source AI now has a definition. This it what it means and why it's still tricky. *EuroNews*. https://www.euronews.com/next/2024/08/23/open-source-ai-now-has-a-definition-this-it-what-it-means-and-why-its-still-tricky